\documentclass[aps,prb,twocolumn,showpacs,superscriptaddress]{revtex4}
\usepackage{eurosym}
\usepackage{graphicx}
\usepackage{amssymb}
\usepackage{amsmath}
\usepackage{color}

\begin{document}

\title{Sweeping reciprocal vortex lattice across the Fermi surface: A new
magnetoquantum oscillations effect in the superconducting state}
\author{T. Maniv}
\altaffiliation{e-mail:maniv@tx.technion.ac.il}
\affiliation{Schulich Faculty of Chemistry, Technion-Israel Institute of Technology,
Haifa 32000, Israel}
\author{V. Zhuravlev}
\affiliation{Schulich Faculty of Chemistry, Technion-Israel Institute of Technology,
Haifa 32000, Israel}
\affiliation{Physics Department, Ort Braude College, P.O. Box 78, 21982 Karmiel, Israel }
\date{\today }

\begin{abstract}
It is shown that coherent scatterings by an ordered vortex lattice are
critically enhanced for quasi particles moving in cyclotron orbits on the
Fermi surface through vortex core regions, thus generating significant
quasi-periodic oscillating contributions to the SC free energy as a
function of the inverse magnetic field. The mean frequency of the
oscillation provides a fingerprint of the vortex lattice geometry.
Vortex-lattice disorder, tends to suppress this oscillatory component.
\end{abstract}

\maketitle

Many reports on observation of de Haas van Alphen (dHvA) or Shubnikov-de
Haas (SdH) oscillations in the mixed (vortex) states of strongly type-II
superconductors, have shown additional damping of the oscillations'
amplitude in the vortex state with respect to the normal state signal (see,
e.g.\cite{Janssen98}-\cite{Bergk09}). In all these experiments the
appearance of the extra damping has been found structureless, reflecting
only some average effect of the vortex matter on the fermionic quasi
particles. \ Extracting valuable information about the vortex state from
these data requires development of a comprehensive quantitative theory of
pure strongly type-II superconductors at high magnetic fields and low
temperatures, a theory which does not exist today even within the
conventional BCS framework \cite{Maniv01}. \ Recently, using a theoretical
approach based on an exact perturbative Gorkov-Ginzburg-Landau theory \cite%
{ZM-PRB12}, it was found, however, that on certain cyclotron orbits
fermionic quasi particles (QPs) are singularly coupled to, and coherently
scattered by the vortex lattice, resulting in a new type of magnetoquantum
oscillations superimposed on the usual dHvA oscillations. \ In the present
paper we analyze characteristic features of the discovered oscillatory
effect and discuss their experimental feasibility. The kinematical condition
controlling the predicted effect is shown, by means of a simple model, to be
operative beyond the limitations of the perturbation theory.

Following the theory presented in Ref.\cite{ZM-PRB12} we consider a 2D
strongly type-II (neglecting the effect of SC screening currents)
superconductor in a perpendicular uniform magnetic field $\mathbf{H=}H%
\widehat{z}$. Generalization to isotropic 3D systems is rather
straightforward. It is assumed that the superconductor can be described by
means of BCS-Hamiltonian for the usual singlet $s$-wave electron pairing
(electron spin is neglected for the sake of simplicity). Within mean-field
approximation (but not in a fully self-consistent manner) the order
parameter is described by a general vortex lattice state, $\Delta (\mathbf{r}%
)=\left( \frac{2\pi }{a_{x}^{2}}\right) ^{1/4}\Delta _{0}\varphi _{0}(%
\mathbf{r})$, written in terms of a discrete set of ground-state Landau
orbitals (in symmetric gauge): \newline
$\varphi _{0}(x,y)=e^{ixy}\sum_{n}e^{-i\theta n^{2}+iq_{n}x-(y+q_{n}/2)^{2}}$%
, where $q_{n}=\frac{2\pi }{a_{x}}n=q_{0}n$ , $n=0,\pm 1,\pm 2,...$ with the
lattice spacing $a_{x}$ along the $x$ - axis and \textit{\ }$a_{x}^{2}=\pi /%
\sqrt{1-\left( \theta /\pi \right) ^{2}}$\textit{. }For the Abrikosov
triangular and square lattices:\textit{\ }$\theta =\pi /2$\textit{,}and%
\textit{\ }$\theta =0$ respectively\textit{. }We use dimensionless space
coordinates measured in units of the magnetic length, $a_{H}=\sqrt{c\hbar /eH%
}$. The amplitude of the order parameter, $\Delta _{0}^{2}=S^{-1}\int d^{2}%
\mathbf{r}_{i}\left\vert \Delta (\mathbf{r}_{i})\right\vert ^{2},$ where%
\textit{\ }$S=\pi N$\textit{\ }and\textit{\ }$N$\textit{\ }is the number of
vortices, is treated as a variational parameter with respect to the SC
thermodynamic potential (TP) $\Omega _{sc}\left( \Delta _{0}\right) $, which
can be written as a Taylor expansion in $\Delta _{0}$: 
\begin{equation}
\Omega _{sc}\left( \Delta _{0}\right) =S\left( \frac{\Delta _{0}^{2}}{g_{BCS}%
}\right) +\sum\limits_{n=1}\frac{\left( -1\right) ^{n}}{n}\Omega _{2n}\left(
\Delta _{0}\right)
\end{equation}%
where $\Omega _{2n}\left( \Delta _{0}\right) =Nk_{B}T\left( \frac{\Delta _{0}%
}{\hbar \omega _{c}}\right) ^{2n}I_{2n}$, $g_{BCS}$ is the BCS coupling
constant and$\ \omega _{c}=eH/mc$. \ The quartic term, $I_{4}$ , is the
leading contribution to the SC free energy influenced by the vortex
distribution\cite{Maniv01}. \ An exact expression derived in \cite{ZM-PRB12}
for a given Matzubara frequency $\omega _{\nu }=\pi k_{B}T\left( 2\nu
+1\right) /\hbar $, reads:

\begin{widetext}

\begin{equation}
I_{4}=\int_{0}^{\infty }d\tau _{1}d\tau _{2}d\tau _{3}d\tau _{4}e^{-\varpi
_{\nu }\left( \tau _{1}+\tau _{2}+\tau _{3}+\tau _{4}\right) -in_{F}\left(
\tau _{1}-\tau _{2}+\tau _{3}-\tau _{4}\right) }\frac{\beta \left( \gamma
\right) }{\alpha _{1}+\alpha _{2}+\alpha _{3}+\alpha _{4}}  \label{I4-Gen}
\end{equation}%
where $\varpi _{\nu }=\omega _{\nu }/\omega _{c}$ $\ $and$\
n_{F}=E_{F}/\hbar \omega _{c}-1/2$. \ Here \ $\beta \left( \gamma \right)
=\sum_{\mathbf{G}}\beta _{\mathbf{G}}\left( \gamma \right) $ is a function
of the 4-electron variable $\gamma =\frac{\alpha _{2}\alpha _{4}-\alpha
_{1}\alpha _{3}}{\alpha _{1}+\alpha _{2}+\alpha _{3}+\alpha _{4}}$, $\
\alpha _{j}=1-e^{i\varepsilon _{j}\tau _{j}}$, $\ \varepsilon _{j}=\left(
-1\right) ^{j+1}$ , \ and:

\begin{equation}
\beta _{\mathbf{G}}\left( \gamma \right) =\frac{1}{2}\left\{ \frac{1}{\left(
1+\gamma \right) }\exp \left[ -\left( \frac{1-\gamma }{1+\gamma }\right)
\left\vert \mathbf{G}\right\vert ^{2}\right] +\frac{1}{\left( 1-\gamma
\right) }\exp \left[ -\left( \frac{1+\gamma }{1-\gamma }\right) \left\vert 
\mathbf{G}\right\vert ^{2}\right] \right\}  \label{beta_G}
\end{equation}

\begin{figure}[tbp]
\begin{center}
\includegraphics[scale=1.5]{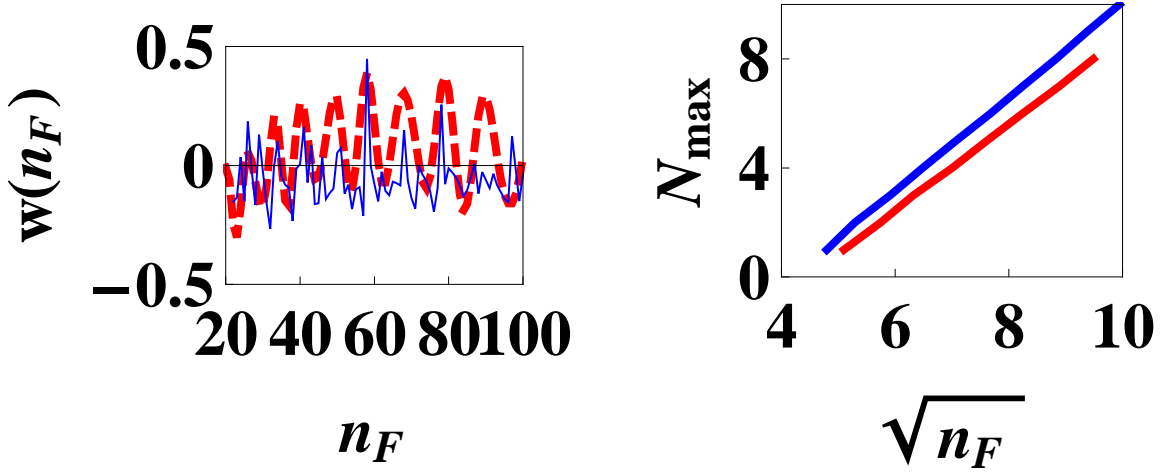}
\end{center}
\caption{(color online): Left panel: The "erratic" function $w\left(
n_{F}\right) $ (solid blue line) plotted together with the characteristic
smooth function $\Xi \left( n_{F}\right) $ (dashed red line) in the interval 
$20\leq n_{F}\leq 100$. Right panel: Number of maxima, $N_{\max }\left( 
\protect\sqrt{n_{F}}\right) $, of the characteristic function $\Xi \left(
n_{F}\right) $ for a square vortex lattice (red line) and for a hexagonal
vortex lattice (blue line). }
\end{figure}

where $\mathbf{G}$ is a reciprocal vortex lattice vector. \ The structure
function $\beta \left( \gamma \right) $ controls the coupling between the
two pairs of electrons involved and the vortex lattice, where $\gamma $ is a
periodic function of the individual electronic "time" variables $\tau _{j}$,
which reflects the underlying electron cyclotron motions. \ Note that $\beta
\left( \gamma =0\right) $ is the well known Abrikosov parameter $\beta _{A}$
for an arbitrary vortex lattice geometry \cite{Maniv01}. The most remarkable
feature of $\beta \left( \gamma \right) $ seen in Eq.(\ref{beta_G}) is
associated with the dual singular points at $\gamma \rightarrow \pm 1$,
where $\tau _{j}$, approach, respectively: $\tau _{1}=\tau _{3}\rightarrow
0,\tau _{2}\rightarrow n\pi -\tau ,\tau _{4}\rightarrow n\pi +\tau $ , or: $%
\tau _{1}\rightarrow n\pi -\tau ,\tau _{3}\rightarrow n\pi +\tau ,\tau
_{2}=\tau _{4}\rightarrow 0$ , with $\tau $ being an arbitrary real number
in the interval: $-\pi \leq \tau \leq \pi $, and $n=1,2,...$. The electrons
at such highly correlated pairs of cyclotron orbits are singularly scattered
by the vortex lattice, yielding only purely harmonic contributions to the SC
free energy in the dHvA frequency $F=n_{F}H$ since under these conditions: $%
e^{-in_{F}\left( \tau _{1}-\tau _{2}+\tau _{3}-\tau _{4}\right) }\rightarrow
e^{-2\pi inn_{F}}$. \ The pair of electrons whose positions on the singular
cyclotron orbit coincide (i.e. $\left( 1,3\right) $ at $\gamma \rightarrow 1$
and $\left( 2,4\right) $ at $\gamma \rightarrow -1$ )\ undergo local mutual
scattering and so exchange many $G$-vectors through the vortex lattice
during the scattering process, while those electrons moving coherently on a
large cyclotron orbit in opposite directions ( i.e. $\left( 1,3\right) $ at $%
\gamma \rightarrow -1$ and $\left( 2,4\right) $ at $\gamma \rightarrow 1$)
are mutually scattered over a long range of the vortex lattice, and so do
not exchange momentum. Thus, the leading purely harmonic contributions to%
\emph{\ }$\beta \left( \gamma \right) $\emph{\ }are: $\beta _{\mathbf{G=0}%
}\left( \gamma \right) =\frac{1}{2}\left( \frac{1}{1+\gamma }+\frac{1}{%
1-\gamma }\right) $, and $\int \beta _{\mathbf{G}}\left( \gamma \right)
d^{2}G$, which yields: $\frac{1}{2}\frac{1}{1-\gamma }$ for $\gamma
\rightarrow 1$ , and $\frac{1}{1+\gamma }$ for $\gamma \rightarrow -1$.

Significant deviations from the harmonic part are developed slightly away
from the singular points. They can be estimated by expanding $\left(
1-\gamma \right) /\left( 1+\gamma \right) $ about the singular points, e.g.
for $\gamma =1$: $\left( 1-\gamma \right) /\left( 1+\gamma \right) \simeq -i%
\widetilde{\xi }_{2}/4+\left( 4\widetilde{\xi }_{1}^{2}+\widetilde{\xi }%
_{3}^{2}\right) /16$ , with $\widetilde{\xi }_{1}=\left( \tau _{1}+\tau
_{2}+\tau _{3}+\tau _{4}\right) /4-\pi /2$ , $\widetilde{\xi }_{2}=\left(
\tau _{1}-\tau _{2}+\tau _{3}-\tau _{4}\right) /2+\pi $ , $\widetilde{\xi }%
_{3}=\tau _{1}-\tau _{3}$ , and carrying out the $\tau _{j}$-integrations. \
Focusing, for simplicity, on the deviations from the first harmonic it is
found that: $I_{4}\rightarrow \mathbf{Re}e^{2\pi in_{F}}e^{-2\pi \varpi
_{\nu }}\int_{0}^{\infty }d\widetilde{\xi }_{1}e^{-4\varpi _{\nu }\widetilde{%
\xi }_{1}}I_{3}\left( \widetilde{\xi }_{1}\right) $, where:

\begin{eqnarray}
I_{3}\left( \widetilde{\xi }_{1}\right) &=&2\mathbf{Re}\sum_{\mathbf{G}%
}\int_{-2\widetilde{\xi }_{1}}^{2\widetilde{\xi }_{1}}\left( 2\widetilde{\xi 
}_{1}-\widetilde{\xi }_{2}\right) d\widetilde{\xi }_{2}\exp \left\{ i%
\widetilde{\xi }_{2}\left[ \frac{1}{4}\left\vert \mathbf{G}\right\vert
^{2}-2n_{F}\right] \right\}  \notag \\
&&\times \int_{-\left( 2\widetilde{\xi }_{1}+\widetilde{\xi }_{2}\right) }^{2%
\widetilde{\xi }_{1}+\widetilde{\xi }_{2}}d\widetilde{\xi }_{3}\exp \left[
-\left( \widetilde{\xi }_{1}^{2}+\frac{1}{4}\widetilde{\xi }_{3}^{2}\right) 
\frac{1}{4}\left\vert \mathbf{G}\right\vert ^{2}\right]  \label{I3}
\end{eqnarray}

\begin{figure}[tbp]
\begin{center}
\includegraphics[scale=1]{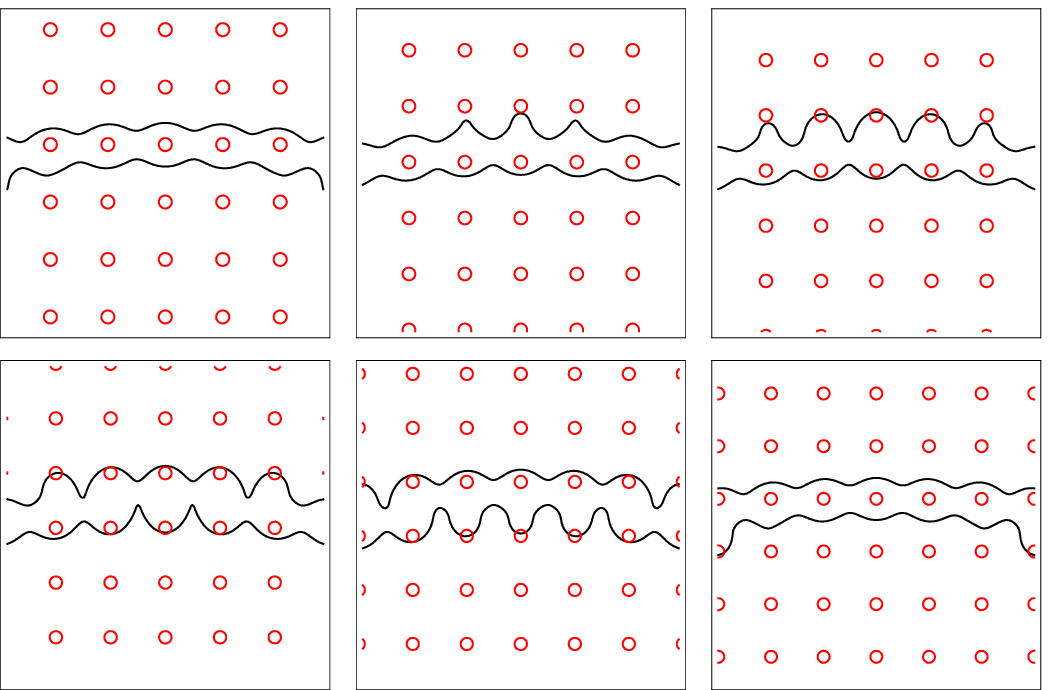}
\end{center}
\caption{(color online): A square lattice of vortex cores (small red
circles), shown sweeping across a pair of classical orbits occupied by a
quasi particle--quasi hole pair (upper and lower solid curves respectively)
in a magnetic field whose magnitude increases successively from the upper
left panel through the lower right panel. Note that in the ground state the
occupied (low-energy branch) cyclotron orbit is locally distorted
paramagnetically in passing through vortex core regions, wherease both
orbital branches of a quasi particle-hole excitation are distorted
diamagnetically.}
\end{figure}

\end{widetext}

It is clear that the dominant contributions to the integral in Eq.(\ref{I3})
originate from reciprocal lattice vectors satisfying: $\left\vert \mathbf{G}%
\right\vert \approx 2\sqrt{2n_{F}}$, namely having length close to the Fermi
surface diameter. Furthermore, due to the large values of $n_{F}$ and the
discrete nature of $\mathbf{G}$ (which are measured in units of $a_{H}^{-1}$%
), the integration over $\xi _{2}$ yields "erratically" oscillating,
quasi-periodic function of $\sqrt{n_{F}}$, superimposed on the usual dHvA
oscillation, expressed by the factor $e^{2\pi in_{F}}$ of $I_{4}$ in our
simple spherical Fermi surface model. \ As will be illustrated below, the
corresponding oscillating envelop function contains vortex-lattice
structural information which can be extracted from experiment by an
appropriate data analysis (see Fig.1).

The final result for the first dHvA harmonic of the SC TP, up to fourth
order, can be written in the form:%
\begin{equation}
\Omega _{sc}^{\left( 1h\right) }/\Omega _{n}^{\left( 1h\right) }\simeq 1-%
\frac{\pi ^{3/2}}{\sqrt{n_{F}}}\left\vert \frac{\Delta _{0}}{\hbar \omega
_{c}}\right\vert ^{2}+\frac{1}{2}\left( 1+w\left( n_{F}\right) \right) \frac{%
\pi ^{3}}{n_{F}}\left\vert \frac{\Delta _{0}}{\hbar \omega _{c}}\right\vert
^{4}-...  \label{Omega-1h}
\end{equation}%
where $\Omega _{n}^{\left( 1h\right) }$ is the corresponding normal state
free energy extrapolated to the vortex state, and $w\left( n_{F}\right) $,
shown in Fig.1,\ represents the "erratically" oscillating component.

The influence of vortex-lattice disorder on the SC free energy in the white
noise limit can be determined by invoking the general expansion of the state
function $\varphi _{0}(x,y)$ in terms of Landau orbitals wave functions with
randomly selected coefficients. Averaging $\beta \left( \gamma \right) $
over realizations of these coefficients it can be easily shown that: $%
\left\langle \beta \left( \gamma \right) \right\rangle \rightarrow \frac{1}{%
1-\gamma }+\frac{1}{1+\gamma }$, which is just the purely harmonic component
of $\beta \left( \gamma \right) $. \ In this limiting case, only incoherent
scattering processes by the vortex matter contribute to the SC TP, and the
final result, up to fourth order, is obtained from Eq.(\ref{Omega-1h}) by
taking $w\left( n_{F}\right) \rightarrow 0$, i.e. very close to the well
known Maki expression\cite{Maki91}, as expanded to the same power in $\Delta
_{0}$.

It is interesting to note that the kinematical condition, $\left\vert 
\mathbf{G}\right\vert =2\sqrt{2n_{F}}$ , is equivalent to the real space
condition, $\left\vert \mathbf{R}\right\vert =2\sqrt{2n_{F}}$ (where $%
\mathbf{R}$ is a vortex-lattice vector measured in units $a_{H}$\ ), which
is just the condition for the cyclotron orbit at the Fermi energy to pass
through a vortex core. The "erratic" oscillations are closely related to
such cyclotron orbits since the latter are strongly distorted during their
passage near a vortex core \cite{Maniv01}. The influence of a "sweeping"
vortex lattice (generated by a sweeping magnetic field) crossing a
quasi-particle quasi-hole pair of cyclotron orbits may be studied
qualitatively by considering a highly simplified model of a classical
charged particle moving in two dimensions under a perpendicular magnetic
field $\mathbf{H}=H\widehat{z}$ and in the presence of a "pair-potential", $%
\left\vert \Delta \left( \mathbf{r}\right) \right\vert $ , with the
Bogoliubov dispersion relation $\epsilon \left( \mathbf{k}\right) =\pm \sqrt{%
\xi _{k}^{2}+\left\vert \Delta \left( \mathbf{r}\right) \right\vert ^{2}}$,
where$\ \xi _{k}=\frac{\hbar ^{2}k^{2}}{2m^{\ast }}-E_{F}$ \cite{Maniv01}.
An important advantage of this model, despite its obvious shortcomings, is
that it can be implemented without invoking perturbation theory with respect
to $\Delta \left( \mathbf{r}\right) $. \ The corresponding coupled
Lorentz-Bloch velocity equations:%
\begin{equation*}
\frac{d\mathbf{k}}{dt}=-\left[ \frac{d\mathbf{r}}{dt}\times \widehat{z}%
\right] \ ;\ \ \ \ \ \frac{d\mathbf{r}}{dt}=\frac{1}{\hbar }\frac{d}{d%
\mathbf{k}}\epsilon \left( \mathbf{k}\right) \ .\ \ 
\end{equation*}%
can be easily integrated to yield the trajectory equation: 
\begin{equation*}
\left[ r^{2}-\left( k_{F}/h\right) ^{2}\right] ^{2}+4\left\vert \frac{\Delta
\left( r,\phi \right) }{\hbar \omega _{c2}h^{2}}\right\vert ^{2}=r_{0}^{4} 
\end{equation*}%
where $h=H/H_{c2}$ and $r_{0}$ is an integral of motion determined by
initial conditions. Obviously, there are two types of solutions for energies
above and below the Fermi energy, corresponding to the Bogoliubove
particle-hole pair. Examples of classically allowed, distorted cyclotron
trajectories, which correspond to real solutions $r\left( \phi \right) $ ,
are depicted in Fig.2. They are restricted to QP energies above the
threshold $r_{0}^{2}=2\max \left\vert \Delta \left( r,\phi \right)
\right\vert /\hbar \omega _{c2}h^{2}$. \ Below this threshold there are
complex solutions corresponding to tunneling of QP orbits between vortex
cores (not shown). For increasing magnetic field strength $h$ at a fixed QP
energy, the lattice of vortex cores, shown in Fig.2 successively from the
upper-left panel through the lower-right panel, sweeps across the QP
cyclotron orbits, which undergo a strong local distortion upon crossing a
vortex core. The appearances of such local distortions are closely related
to the "erratic" parmagnetic-diamagnetic oscillations of the function $%
w\left( n_{F}\right) $\ shown in Fig.1. \ A characteristic feature of $%
w\left( n_{F}\right) $ to be exploited for distinguishing between different
vortex-lattice structures can be conveniently defined in terms of the smooth
function $\Xi \left( n_{F}\right) =a\sum_{\mathbf{G}}\exp \left[ -\frac{1}{4}%
\left\vert \mathbf{G}\right\vert ^{2}\left( 1-8n_{F}/\left\vert \mathbf{G}%
\right\vert ^{2}\right) ^{2}\right] -b$ (see Fig.1, left panel), derived
with the help of Eq.(\ref{I3}) by performing the integral $\int_{0}^{\infty
}d\widetilde{\xi }_{1}e^{-4\varpi _{\nu }\widetilde{\xi }_{1}}I_{3}\left( 
\widetilde{\xi }_{1}\right) $ in the stationary phase approximation. Here $a$
and $b$ are adjustable parameters. The number of maxima, $N_{\max }\left( 
\sqrt{n_{F}}\right) $ of $\Xi \left( n_{F}\right) $ appearing bellow a given
value of $n_{F}$ offers the desired characteristic feature (see Fig.1, right
panel). Its slop is a measure of the rate at which vortex cores in Fig.2
cross the cyclotron orbit at the Fermi energy, and so uniquely characterizes
the point symmetry of the underlying vortex lattice.

\begin{acknowledgements}
This research was supported by Posnansky Research fund in superconductivity, and by EuroMagNET under the EU contract No.\ 228043. V.Z. acknowledges the support of the Israel Science Foundation, Grant No. 249/10 
\end{acknowledgements}

\end{document}